# Binding events through the mutual synchronization of spintronic nano-neurons

Miguel Romera[1,2,3,†], Philippe Talatchian[1,†], Sumito Tsunegi[4], Kay Yakushiji[4], Akio Fukushima[4], Hitoshi Kubota[4], Shinji Yuasa[4], Vincent Cros[1], Paolo Bortolotti[1], Maxence Ernoult[1,5], Damien Querlioz[5], Julie Grollier[1*]

[1] - Unité Mixte de Physique, CNRS, Thales, Université Paris-Saclay, 91767 Palaiseau, France

[2] - Grupo de Física de Materiales Complejos, Dpto. Física de Materiales, Universidad Complutense de Madrid, 28040 Madrid, Spain

[3] - Unidad Asociada UCM/CSIC, Laboratorio de Heteroestructuras con Aplicación en Espintrónica, 28049 Madrid, Spain

[4] - National Institute of Advanced Industrial Science and Technology (AIST), Spintronics Research Center, Tsukuba, Ibaraki 305-8568, Japan

[5] - Université Paris-Saclay, CNRS, Centre de Nanosciences et de Nanotechnologies, 91120 Palaiseau, France

[†] These two authors have equally contributed to the work

[*] julie.grollier@cnrs-thales.fr

**The brain naturally binds events from different sources in unique concepts. It is hypothesized that this process occurs through the transient mutual synchronization of neurons located in different regions of the brain when the stimulus is presented**[1–4]**. This mechanism of 'binding through synchronization' can be directly implemented in neural networks composed of coupled oscillators**[5]**. To do so, the oscillators must be able to mutually synchronize for the range of inputs corresponding to a single class, and otherwise remain desynchronized. Here we show that the outstanding ability of spintronic nano-oscillators to mutually synchronize**[6–9] **and the possibility to precisely control the occurrence of mutual synchronization by tuning the oscillator frequencies over wide ranges allows pattern recognition. We demonstrate experimentally on a simple task that three spintronic nano-oscillators can bind consecutive events and thus recognize and distinguish temporal sequences. This work is a step forward in the construction of neural networks that exploit the non-linear dynamic properties of their components to perform brain-inspired computations.**

Spintronic oscillators are nanoscale devices called magnetic tunnel junctions which have the potential to be integrated by hundreds of millions in electronic chips[10]. The microwave voltages that they produce have varying amplitude and frequency in response to direct current inputs. Their non-linear dynamical properties are rich and tunable, and can be leveraged to imitate different features of biological neurons, which makes them particularly promising for neuromorphic computing[11–15]. The transient dynamics of a single spintronic nano-oscillator has been used to implement reservoir computing, achieving state-of-the-art results on a simple spoken digit recognition task[16,17]. Four spintronic nano-oscillators have been trained to classify spoken vowels by phase locking their oscillations to the strong input signals produced by external microwave sources[18].

It is now essential to demonstrate that the mutual synchronization of spintronic nano-oscillators can be exploited for computing. Larger hardware networks of oscillators can be built if the oscillators directly influence each other and synchronize through the weak signals that they emit, without the need of high power-consumption amplification stages. The latter is possible with spin-torque nano-oscillators due to their outstanding synchronization ranges, enhanced by factors of typically ten compared to Kuramoto model-like phase oscillators due to the coupling between their amplitude and phase[6–9,19–21]. In addition, more complex tasks can be achieved by exploiting the rich interactions that emerge in assemblies of mutually coupled oscillators rather than using phase-locking to external signals[22,23].

A primary source of inspiration to move in this direction is neuroscience, which shows that in the brain, vast groups of neurons mutually synchronize in response to external or internal stimuli, giving rise to strong oscillatory signals[1–3]. This process is often hypothesized to enable spatiotemporal integration of stimuli, a mechanism called 'binding through synchrony'[4]. It can be used for pattern recognition in oscillatory neural networks if mutual synchronization can be controlled and tuned to achieve the desired task[5].

In this work we show that spintronic nano-oscillators can recognize temporal patterns through their mutual synchronization by binding together consecutive events in time. We implement a hardware neural network based on these principles with three spintronic nano-oscillators. We demonstrate that it recognizes sequences of spikes from a neuroscience-inspired database with a success rate of 94%, approaching the success rate of 96% achieved by identical and noiseless oscillators. We show that these high recognition rates stem from the possibility to precisely control the mutual synchronization of spintronic nano-oscillators by varying their frequency over large direct current input ranges.

Our experiment exploits the coupling that occurs naturally when hardware spintronic nano-oscillators are electrically connected to synchronize them with each other[9]. The set-up is shown in Fig. 1a (details on samples and set-up are given in Methods). An important feature of these nano-oscillators is that their frequency can be individually and easily controlled by varying the direct current through each oscillator. When their frequencies are well separated, the three oscillators emit microwave signals independently, and the spectrum analyzer at the output of the set-up displays three peaks (Fig. 1b). The propagation of these microwave emissions in the line creates a coupling between the connected oscillators[9]. When the frequencies of the oscillators are brought closer together, within the mutual locking range – here of 5 MHz –, they synchronize, which results in the single peak of Fig. 1c. Its power is significantly higher than the sum of the individual emitted powers of the three disconnected oscillators measured with the same bias conditions (Extended Data Fig. 1 and 2). This distinctive feature is due to the phase coherence between oscillators characteristic of the synchronized state (see Methods)[19,20].

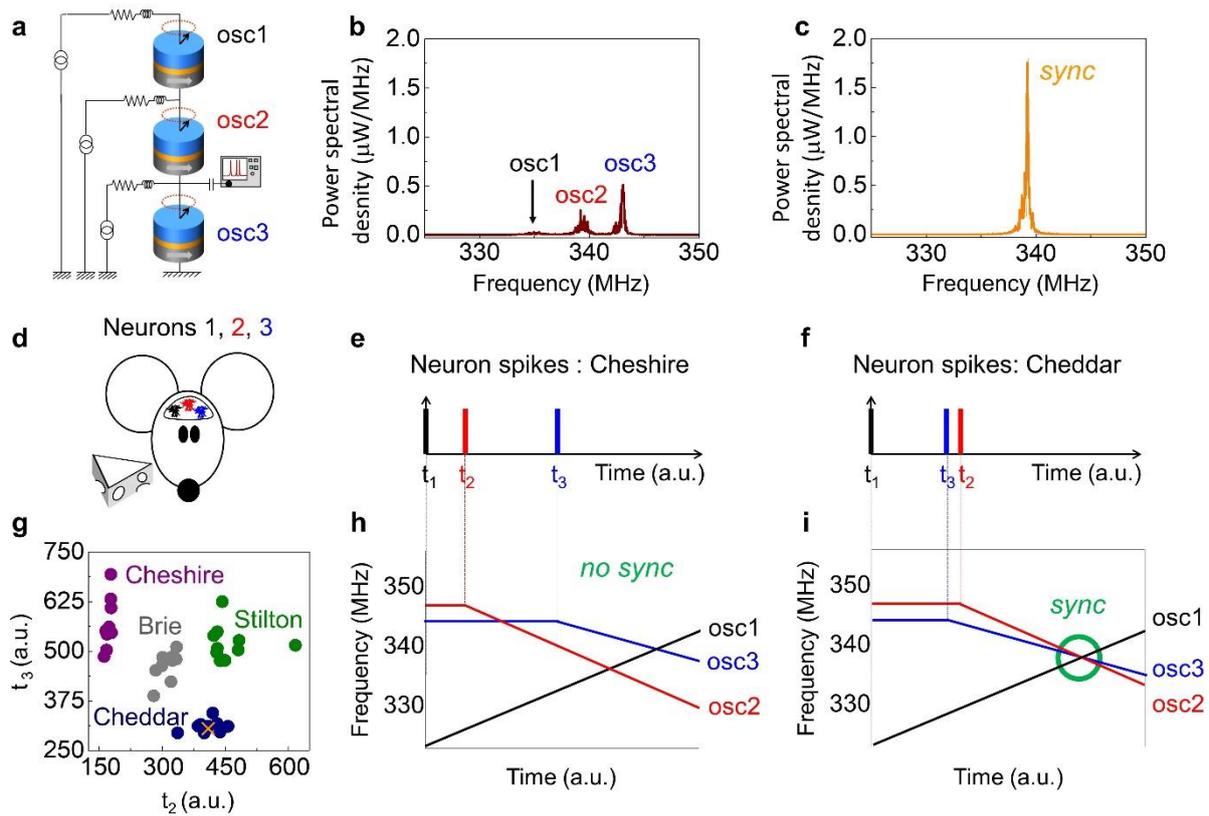

**Figure 1: Binding temporal sequences through synchrony.** (a) Schematic of the experimental set-up with three spin torque nano-oscillators electrically connected in series and coupled through the microwave currents they emit. (b, c) Microwave output emitted by the network of three coupled oscillators when they are not (b) and when they are (c) synchronized. (d) Schematic of the fictitious mouse to which four different categories of cheese are presented. Each category generates different activities in the three neurons of the mouse brain. (e, f) Sequence example of neuron spikes in the mouse brain in presence of a piece of Cheshire (e) and Cheddar (f) respectively. (g) Inputs applied to the network, represented as the time of spikes for neuron 2 and 3. The spike of neuron 1 is set as origin of the sequence and taken as zero. Each color corresponds to a different cheese category, and each data point corresponds to a different piece of cheese. (h, i) Ramps of current generated in the network upon application of the input spike sequences described in (e) and (f), corresponding to the presentation to the mouse of a piece of Cheshire and Cheddar respectively, when the network is trained to recognize Cheddar.

We leverage this mutual synchronization phenomenon to recognize temporal sequences with a hardware neural network of three spintronic oscillators. We consider a fictitious task, in which a mouse is presented with four different categories of cheese (Fig. 1d). Each kind of cheese generates different neuron activities in the mouse brain. The temporal sequences to be classified are composed of three spikes, each recorded from three different neurons of the fictitious mouse brain. Fig. 1e and Fig. 1f illustrate sequences of spikes when the mouse is in presence of Cheshire and Cheddar cheese. By convention, the time of the first spike is taken as zero and the times of spikes for neurons 2 and 3 are shown in Fig. 1g for the whole database. The different colors correspond to different categories of cheese. Ten different samples of cheese per category are presented to the mouse, giving rise to the variability within each category. The goal of the task is to infer which type of cheese is presented to the mouse by analyzing the recorded sequence of spikes.

For this purpose, we use a mechanism initially proposed by Hopfield et al. to bind temporal features through neural synchronization[4]. A network is composed of as many neurons as there are spikes in the input sequence, and is trained to recognize a single category of input. Our dataset features three spikes, therefore we use three spintronic oscillators tuned to recognize a given kind of cheese, e.g. Cheddar. Each spike triggers a current ramp in the associated neuron. Fig. 1h is a schematic illustrating the behavior of the network activated by a spike sequence that it has not been designed to recognize, e.g. Cheshire cheese. In that case, the ramps do not intersect at any time. On the contrary when the oscillator network is activated with the spike sequence for Cheddar cheese that it has been designed to recognize, the different ramps intersect at a specific time, as illustrated in Fig. 1i. The neurons therefore transiently mutually synchronize and give rise to a large output signal as in Fig. 1c signaling that they have bonded the events together and identified the sequence as meaningful.

In this framework, training the network means finding parameters for the ramps leading to mutual synchronization of the three oscillators for all the points of the database corresponding to the Cheddar category, even if the corresponding spike times are scattered in time. Experimentally, we use the center of the 'Cheddar' data points cloud, pinpointed with an orange cross in Fig. 1g, as a target to calibrate the network, and we convert the arbitrary units of Fig. 1g to seconds. The ramps of current in oscillator 2 and oscillator 3 are therefore triggered with delays of 412 s and 308 s respectively for this calibration point. We choose the oscillators initial frequencies and the slope of the ramps of applied currents in such a way that the application of this sequence of spikes eventually leads to a set of currents $I_{Synch}$ = (6.8 mA, 6.2 mA, 6 mA) for which the oscillators synchronize (details on the calibration procedure are given in Methods). Once this calibration is done, when a new input is presented to the network, the selected initial current conditions ($I_{osc1}^0$, $I_{osc2}^0$, $I_{osc3}^0$) = (4.9 mA, 7.51 mA, 5.15 mA) and the slopes of the ramps of current ($dI_{osc1}/dt$, $dI_{osc2}/dt$, $dI_{osc3}/dt$) = (2.5 µA/s, -3.75 µA/s, 1.875 µA/s) are maintained.

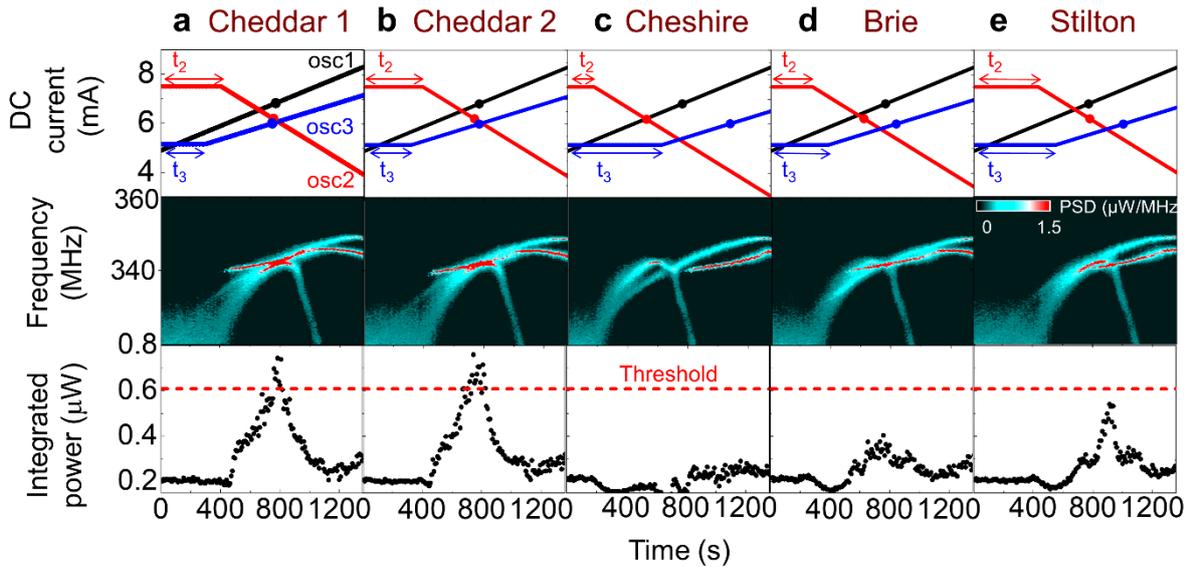

**Fig 2. Three hardware spintronic nano-oscillators recognize cheddar.** Response of the oscillator network trained to recognize Cheddar to spikes sequences generated in the mouse brain when a piece of Cheddar (a,b) or a different cheese (c-e) is presented. The spike sequences generate ramps of currents (top) which translate into variations of the oscillators frequencies (middle) and the network total emitted power (bottom). If a piece of Cheddar is presented (a,b), the ramps of current and their associated variations of frequencies lead to transient mutual synchronization of the three oscillators. This translates into an enhancement of the network total emitted power above the threshold shown in red dotted line (a-b, bottom) meaning recognition. If a different cheese is presented (c-e), the ramps of current do not give rise to the mutual synchronization of the three oscillators and the total emitted power remain well below the threshold (c-e, bottom), meaning that the network distinguishes that the cheese presented is not Cheddar.

Fig. 2a shows the measured oscillator responses to the sequence 'Cheddar 1' = (400 s, 294 s), which is the first entry in the class 'Cheddar'. The corresponding trained ramps of currents are triggered at $t_1$ = 0, $t_2$ = 400 s and $t_3$ = 294 s in oscillator 1, 2 and 3 respectively. The three large dots in the top panel of Fig. 2 highlight the set of currents $I_{synch}$ that we have used to calibrate the network, at which the oscillators synchronize. As can be seen in Fig. 2a, after ~760 seconds the direct currents flowing through the oscillators reach values approaching $I_{synch}$. At this point, the frequencies of all three oscillators become near-identical (Fig. 2a-middle) and the total emitted power peaks at $P_{max}$ = 0.743 µW. To know if the oscillators have recognized the category Cheddar, we need to assess if they have transiently synchronized or not. For this purpose, we can compare $P_{max}(\mathbf{I})$ to the sum of the powers of the three independent oscillators for the same current conditions $P_{unsync}(\mathbf{I})$, equal to 0.43 µW at this particular point. Here $P_{max}$ is much larger than $P_{unsync}$, which shows that the oscillators have mutually synchronized and successfully recognized that the piece of cheese belongs to the Cheddar category.

The value of $P_{unsync}$ varies strongly with the currents $\mathbf{I}$ applied to the oscillators which makes it difficult to use it as a general criterion to detect synchronization for all sets of currents in the experiment. In the following we consider that the network recognizes the input pattern if the total emitted power reaches a fixed threshold value of 0.608 µW, independent of the current values (dotted red line in the bottom panel of Fig. 2) and if the frequencies of the three oscillators at that point are all equal (see Methods). Fig.2b shows the network response to another input from the Cheddar class. For this sequence 'Cheddar2' = (392 s, 315 s), the triggered ramps of current never reach $I_{synch}$, but they get close to this value around t = 744 s, where a mutual synchronization event is observed, with the emission of a single peak accompanied with a large increase of the emitted power above the threshold. This input is therefore correctly classified as belonging to the category 'Cheddar'. This example illustrates well the interest of using mutual synchronization to categorize spread data: synchronization occurs when oscillators frequencies are close, but they do not need to be exactly identical (as in Extended Data Fig. 1).

In contrast, Fig. 2c-e show the network response to inputs of the other cheese categories, which characteristic time sequences never lead to sets of currents close to $I_{synch}$. In consequence, the three oscillators do not synchronize and the power emitted by the network remains well below the recognition threshold. In these three cases (Fig. 2c-e) the network interprets correctly that the applied input does not correspond to the class 'Cheddar'.

Table 1 shows the overall classification performances of the network. As can be seen in row 1, the network classifies correctly 9 out of 10 inputs corresponding to the class 'Cheddar'. Moreover, when inputs from other categories ('Cheshire', 'Brie', and 'Stilton') are applied to the system, the network correctly interprets that the input is not a piece of 'Cheddar'. The overall success rate for this category is 97.5%.

| Cheese that the oscillators are trained to detect | Presented cheese (10 datapoints) | | | | Correct response of the network |
|---|---|---|---|---|---|
| | Cheddar | Brie | Cheshire | Stilton | |
| | Number of recognitions (out of 10) | | | | |
| **Cheddar** | **9** | 0 | 0 | 0 | **97.5 %** |
| **Brie** | 0 | **10** | 0 | 3 | **92.5 %** |
| **Cheshire** | 0 | 0 | **7** | 0 | **92.5 %** |
| **Stilton** | 0 | 0 | 0 | **8** | **95 %** |

**Table 1: Recognition rates.** Number of recognitions out of 10 presented samples of each cheese, when the network is trained to classify Cheddar, Brie, Cheshire and Stilton respectively. The last column refers to the percentage of times that the network responds correctly, either because it detects that the input belongs to the category it was trained to recognize or because it interprets correctly that the inputs corresponds to another cheese category.

The same three oscillators can also be trained to recognize the other categories of cheese. For this we modify the initial conditions of the network (the initial currents flowing through each oscillator $I_{osci}^0$, and the slopes of the ramps of current triggered when an input is applied, $dI_{osci}/dt$) in such a way that the oscillator currents reach $I_{synch}$ only when the desired category of input is applied to the system (see Methods). Using these new calibration parameters, we repeat the recognition experiment by applying the same dataset as previously. The results are shown in Table 1: the network recognition rate for 'Brie', 'Cheshire' and 'Stilton' is respectively 92.5, 92.5 and 95%. Overall, the network responds correctly to 94% of the inputs. By comparison, a simulated network of three identical, noiseless oscillators gives a recognition rate of 96% on the same database (see Methods).

The excellent performance of this simple network comes from the match between the requirements of the algorithm and the physical properties of spintronic nano-oscillators. Here the inputs are sequences of events that are largely spread in time and need to be bonded to constitute a single concept. The algorithm does this task in two steps. First, it converts the spread timing of events in the sequence to close-by neuron frequencies by ramping the values of frequencies over wide ranges. The high frequency tunability of spintronic nano-oscillators[24] provides a straightforward hardware implementation of this property. Second the algorithm leverages neuron synchronization to bind these neighboring frequencies into a single concept, here, a cheese category. This synchronization range must be large enough to ensure that events are bonded even if different sequences encoding the same category of cheese are scattered in time. The large synchronization bandwidths accessible to spintronic nano-oscillators[6] reduce the precision requirements on the frequency, and therefore on the direct current steps, needed to achieve ramp convergence.

Complementary Metal Oxide Semiconductor (CMOS) technology-based voltage-controlled oscillators such as ring oscillators can also exhibit such characteristics, but large-area decoupling capacitors are needed to control their synchronization, and digital to analog converters to read their outputs[25]. The total silicon area occupied by a single oscillator is larger than 3x3 µm$^2$,[26] whereas spin-torque nano-oscillators can be scaled below 100x100 nm$^2$. Using small-area oscillators is imperative, as moving beyond toy tasks will require scaling up the system. Our simple network processes inputs composed of three events, but more complex inputs will be based on a larger number of events and will require more oscillators. The original paper of Hopfield uses 40 neurons for recognizing the first ten spoken digits[4]. It has been shown recently that such large numbers of spintronic nano-oscillators can mutually synchronize, by driving them with spin-Hall torques and coupling them strongly through ferromagnetic exchange[27]. Important for scaling to even larger dimensions, the algorithm is very tolerant to device variability: as long as the oscillators synchronize over large bandwidths, current ramps can be found for every input though training[4]. Recognition can potentially be achieved in just a few tens of nanoseconds, as spintronic nano-oscillators can be tuned and synchronized within these timescales[28,29]. Finally, when the oscillators synchronize, they generate an additional direct voltage through an effect called spin-diode, that can be detected and then processed with simple and energy efficient CMOS circuits[18]. This work therefore constitutes a milestone towards the implementation of large scale oscillatory neural networks using the physical properties of spintronic nano-oscillators to compute.

**Acknowledgements**

This work was supported by the European Research Council ERC under Grant bio*SPIN*spired 682955. M.R. acknowledges Spanish MINECO (PGC2018-099422-A-I00) and Comunidad de Madrid (2018-T1/IND-11935).

**Author contributions**

The study was designed by J.G., samples were optimized and fabricated by S.T. and K.Y., experiments were performed by M.R. and P.T., numerical simulations were realized by P.T. and M. E. All authors contributed to analyzing the results and writing the paper.

METHODS

**Samples**

Magnetic tunnel junctions (MTJs) with a structure of buffer/PtMn(15)/Co$_{71}$Fe$_{29}$(2.5)/Ru(0.9)/Co$_{60}$Fe$_{20}$B$_{20}$(1.6)/Co$_{70}$Fe$_{30}$(0.8)/MgO(1)/Fe$_{80}$B$_{20}$(6)/MgO(1)/Ta(8)/Ru(7) (thicknesses in nm) were deposited by ultrahigh-vacuum (UHV) magnetron sputtering. After annealing at 360 °C for one hour, the resistance–area product was $RA \approx 3.6$ Ω µm$^2$. Circular-shaped MTJs with a diameter of about 375 nm were patterned using Ar ion etching and e-beam lithography. The resistance of the devices is close to 40 Ω, and the magneto-resistance ratio is about 100% at room temperature. For the dimensions used here, the FeB layer presents a magnetic vortex as the ground state. In the vortex core (a small region of about 12 nm diameter at remanence for our materials), the magnetization spirals out of plane. Under direct current injection and the action of the spin transfer torques, the core of the vortex steadily gyrates around the center of the dot with a frequency in the range of 150 MHz to 450 MHz for the oscillators we used here.

**Experimental set-up**

Fig. 1a shows a schematic of the experimental set-up with three electrically coupled vortex nano-oscillators. A magnetic field of $\mu_0 H$ = 400 mT is applied perpendicularly to the oscillator layers to get an efficient spin transfer torque acting on the oscillator vortex core. A direct current is injected into each oscillator to induce vortex dynamics, which leads to periodic oscillations of the magnetoresistance, giving rise to an oscillating voltage at the same frequency than the vortex core dynamics. The three oscillators are electrically connected in series by millimetre-long wires. In this configuration, the microwave current generated by each oscillator propagates in the electrical microwave circuit influencing the dynamic and in particular the frequency of the other oscillators through the microwave spin-torques it creates. The oscillators are therefore electrically coupled through the microwave currents they emit, and too far away to be coupled through the magnetic dipolar fields that they radiate. Three direct currents ($I_{DC1}$, $I_{DC2}$, $I_{DC3}$) are supplied to the circuit by three different sources, allowing an independent control of the current flowing through each oscillator. Thus, we can control the frequency of each oscillator independently. The actual current flowing through each spin-torque oscillator is given by $I_{STO1} = I_{DC1}$, $I_{STO2} = I_{DC2} + I_{DC1}$, and $I_{STO3} = I_{DC3} + I_{DC2} + I_{DC1}$, respectively, where $I_{STOi}$ corresponds to the current flowing through the $i$th oscillator. The microwave signal emitted by the coupled system is recorded by a spectrum analyzer.

**Synchronization detection**

When spin torque oscillators mutually synchronize their non-linear magnetization dynamics reaches a new state characterized by the oscillators phases being locked to each other. This stabilizes their oscillations frequencies and reduces the main sources of noise in the magnetization dynamics: the amplitude noise and the phase noise, the latter being particularly disruptive for the oscillations coherence. In consequence, the signature of mutual synchronization state is an emission spectrum which shows a drastic increase of the spectral coherence. This is characterized by an emitted power which is above the sum of the powers emitted by the individual oscillators when they are not coupled[8,9,27].

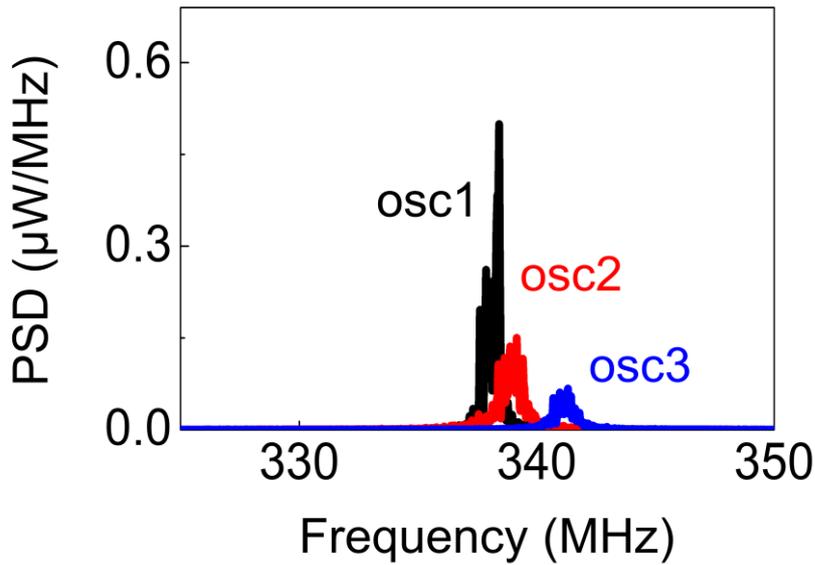

**Extended data Figure 1: Uncoupled oscillator responses.** Microwave output emitted by the three oscillators under the same conditions of field and current of Fig. 1c but when they are not connected to each other.

Extended Data Fig. 1 shows the emitted spectra of the three oscillators under the same conditions of applied current as in Fig. 1c but when they are not connected to each other. As can be seen the frequencies are close to each other but they are not equal. This result illustrates well the interest of using mutual synchronization to categorize spread data: oscillators frequencies do not need to be exactly identical to observe mutual synchronization, as soon as they are closer than the locking range.

Extended Data Fig. 2 shows the oscillators network emitted power (black dots) during the experiment shown in Fig. 3a. The total emitted power reaches a maximum value of $P_{max}(I)= 0.743\mu W$. This value is significantly higher than the sum of the individual emitted powers of the three oscillators (dash red line), which is around $P_{unsync}(I)=0.43\mu W$ at its maxima in this particular experiment. The value of $P_{unsync}$ varies strongly with the currents $I$ applied to the oscillators. Additionally, synchronization events of only two oscillators but at large applied currents (i.e. large emitted power) may translate into large values of the total emitted power. This makes difficult to use a unique general criterion to detect pattern recognition independently of the set of applied currents **I**. Indeed, standard synchronization analysis requires comparing the synchronize state with the emission properties of the non-interacting oscillators under the same conditions for each specific data point within the experiment. Thus, in order to find a unique general criterion for pattern classification independent of the values of applied current **I**, and thus of the current ramps of the particular experiment, we consider recognition when the total emitted power overcomes a threshold value of 0.608 µW and the oscillators have the same frequency at that point. The threshold value has been chosen to minimize misclassification errors, considering the emitted power of the independent oscillators at all possible applied currents within the experimental range.

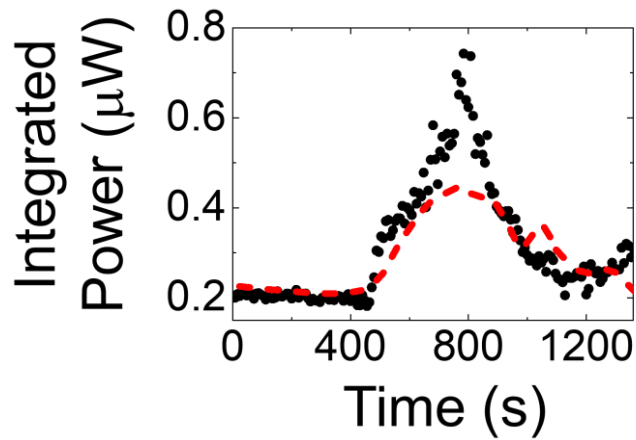

**Extended data Figure 2**. Integrated power of the signal emitted by the three coupled oscillators during the experiment shown in Fig. 2a (black dots), and sum of the integrated power emitted by the three oscillators when they are measured independently under the same conditions (dashed red line).

**Trained initial conditions to classify each category of cheese / Calibration parameters**

The network of oscillators can learn to classify new data. Here, training means finding the parameters of the ramps of current (i.e. initial currents flowing through each oscillator and slopes of the ramps triggered when an input is applied) that lead to mutual synchronization (recognition) when a particular class of input is applied. To do so, we first find conditions of current at which the three oscillators mutually synchronize, $I_{Synch}$ = (6.8 mA, 6.2 mA, 6 mA). Second, we choose for each category a sequence of delays that we want to target. In this study, we used the center of the cloud of each category in Fig. 1g to define these target sequences.

Then we determine three ramps of currents which, upon applying these delays, eventually lead to $I_{Synch}$. To simplify the training, we keep the ramp of oscillator 1 fixed for all categories and vary only the ramps of oscillators 2 and 3. There are many possible initial conditions for oscillators 2 and 3 that eventually lead to $I_{Synch}$. In order to minimize misclassification errors, we choose ramps with the following constraints.

- The slopes of current ramps are chosen so that the frequency step is smaller than the synchronization range. Otherwise, synchronization is not detected which will cause classification errors. Large locking ranges are thus beneficial as they reduce the required precision in terms of frequency or applied current to apply this procedure.

- The signs of the slopes are chosen to maintain the total emitted power in the absence of synchronization as low as possible in the whole range of applied direct currents. This facilitates the synchronization detection by the increase of emitted power.

Once we have chosen slopes satisfying the previous conditions, we set the initial values of applied current that will lead to $I_{Synch}$ upon application of such slopes. For more complex networks, ramp parameters can be found through standard linear regression or gradient descent procedures.

Extended data Fig.3 shows the trained calibration parameters used for each input that the network is trained to classify.

| Cheese that the oscillators are trained to detect | Trained Initial Conditions $I_{STOi}^0$ (mA) $dI_{STOi}/dt$ (µA/s) | | |
|---|---|---|---|
| | STO1 | STO2 | STO3 |
| | $I_{STO1}^0 = 4.9$ | $I_{STO2}^0 = 7.51$ | $I_{STO3}^0 = 5.15$ |

| Cheddar | $dI_{STO1}/dt=+2.5$ | $dI_{STO2}/dt=-3.75$ | $dI_{STO3}/dt=+1.875$ |
|---|---|---|---|
| Brie | $I_{STO1}^0=4.9$ | $I_{STO2}^0=5.07$ | $I_{STO3}^0=7.11$ |
|  | $dI_{STO1}/dt=+2.5$ | $dI_{STO2}/dt=+2.5$ | $dI_{STO3}/dt=-3.75$ |
| Cheshire | $I_{STO1}^0=4.9$ | $I_{STO2}^0=4.82$ | $I_{STO3}^0=7.05$ |
|  | $dI_{STO1}/dt=+2.5$ | $dI_{STO2}/dt=+2.5$ | $dI_{STO3}/dt=-3.45$ |
| Stilton | $I_{STO1}^0=4.9$ | $I_{STO2}^0=5.42$ | $I_{STO3}^0=6.92$ |
|  | $dI_{STO1}/dt=+2.5$ | $dI_{STO2}/dt=+2.5$ | $dI_{STO3}/dt=-3.75$ |

**Extended Data Figure 3.** Trained calibration parameters to classify each category of cheese.

**Simulations with ideal oscillators**

The pattern recognition scheme of the experiment was simulated with a network of three identical noiseless oscillators, using the database of Fig. 1g. The only parameter that differs from one simulated oscillator i from the other one is its applied direct current $I_i$. The simulated oscillators correspond to vortex-based spin-torque oscillators as in the experiment. Their dynamics follows the differential Thiele equation model:

$$\boldsymbol{G}_i \times \frac{d\boldsymbol{X}_i}{dt} - \widehat{D}_i(\boldsymbol{X}_i)\frac{d\boldsymbol{X}_i}{dt} - \frac{\partial W_i(\boldsymbol{X}_i,I_i,I_{com}^{rf})}{\partial \boldsymbol{X}_i} + \boldsymbol{F}_i^{STT}(\boldsymbol{X}_i,I_i,I_{com}^{rf}) = \boldsymbol{0}$$

Here, $\boldsymbol{X}_i = \begin{pmatrix} x_i \\ y_i \end{pmatrix}$ is the vortex core position, $\boldsymbol{G}_i$ is the gyrovector, $\widehat{D}_i$ is the damping, $W_i$ is the potential energy of the vortex, $\boldsymbol{F}_i^{STT}$ is the spin-transfer force, and $I_{com}^{rf}$ is a common microwave current $I_{com}^{rf}$. This model successfully describes experimental results with spin-torque nano-oscillators and can easily be generalized to nonlinear auto-oscillators as van der Pol oscillators[18]. The parameters used for the Thiele equation in the simulation are expressed in Extended Data Fig. 4.

| Parameters | Symbol | Value |
|---|---|---|
| Tunnel magnetoresistance ratio | $TMR$ | 74 % |
| Linear damping | $D$ $(kg\ rad^{-1}s^{-1})$ | $4.28 \times 10^{-15}$ |

| Gyrovector amplitude | $G$ $(kg\ rad^{-1}s^{-1})$ | $2.00 \times 10^{-13}$ |
|---|---|---|
| Slonczewski-like torque efficiency | $a_j\ (kg\ m^2 A^{-1}s^{-2})$ | $3.90 \times 10^{-16}$ |
| Field-like torque efficiency | $b_j$ $(kg\ m^2 A^{-1}s^{-2})$ | $8.44 \times 10^{-23}$ |
| Linear magneto-static confinement | $\kappa_{ms}\ (kg\ s^{-2})$ | $4.05 \times 10^{-4}$ |
| Nonlinear magneto-static confinement | $\kappa'_{ms}\ (kg\ s^{-2})$ | $1.01 \times 10^{-4}$ |
| Linear Oersted field confinement | $\kappa_{Oe}\ (kg\ m^2 A^{-1}s^-)$ | $1.42 \times 10^{-15}$ |
| Nonlinear Oersted field confinement | $\kappa'_{Oe}\ (kg\ m^2 A^{-1}s^-)$ | $-7.12 \times 10^{-15}$ |
| Nonlinear Damping parameter | $\xi$ | 1.6 |

**Extended data Figure 4.** Parameters used to simulate oscillators through the Thiele equation.

A fourth order Runge-Kutta scheme is used to solve simultaneously the three coupled differential Thiele equations corresponding to the three coupled oscillators. The integration time step was set to 0.01 ns. Simulations are achieved at T = 0 K (no thermal noise). As in the experiment, the simulated oscillators are electrically coupled through the sum of their individual microwave alternative current emissions. The expression of these common current $I^{rf}_{com}$ is described as follows[18]:

$$I^{rf}_{com} = \frac{1}{Z_0 + \sum_{i=1}^{3} R_i} \left( \sum_{i=1}^{3} \lambda\, \Delta R_i\, I_i y_i \right)$$

Here $\Delta R_i$ is the mean resistance variation caused by the vortex core gyrotropic motion, $I_i$ is the direct current flowing through the i-th oscillator, $R_i$ is its mean electrical resistance, $Z_0 = 50\ \Omega$ is the load impedance and $\lambda = \frac{2}{3}$.

The main variables extracted from the simulations are the three steady state frequency (f1, f2, f3) of the three oscillators obtained at a given set of direct currents (I1, I2, I3). As in the experiment, a given direct current set (I1, I2, I3) corresponds to a step of the current ramp of the network. In order to extract (f1, f2, f3) from the simulations, first the instantaneous frequency of each oscillator is determined through the simulated cartesian trajectory and velocity of the vortex core over 5 µs. Then, the steady state frequency is computed by evaluating the temporal average of the instantaneous frequency over only the last 60% of the simulated time trace corresponding to 3 μs. Due to the electrical coupling, these steady-state frequencies differ from those obtained in the individual uncoupled case. Depending on the direct current received by each oscillator, their frequencies are pulled and can eventually merge leading to a mutual synchronization. As in experiments, a recognition event corresponds to a mutual synchronization of all the three oscillators to a common frequency. In order to systematically detect this type of events in simulations, we analyze the frequency difference between the three oscillators as follows:

a) If $|f_1 - f_2| \leq f_{th}$ and $|f_2 - f_3| \leq f_{th}$ then the three oscillators are mutually synchronized
b) Otherwise, the three oscillators are considered to be not synchronized all together.

Here $f_{th}$ is a threshold value set to 0.1 MHz. Following criteria a) and b), if the three oscillators remain mutually synchronized for at least three consecutive direct current ramp steps corresponding to at least 13 μs, then the simulated network is considered to be in a recognition state. The initial value $I^0$ and slope value dI/dt of the direct current ramps in simulations are calibrated following the method described in the section "Trained initial conditions to classify each category of cheese / Calibration parameters". Extended data Figure 5 shows the trained calibration parameters used for each input that the network is trained to classify.

| Cheese that the oscillators are trained to detect | Trained Initial Conditions $I_{STOi}^0$ (mA) $dI_{STOi}/dt$ (μA/μs) | | |
|---|---|---|---|
| | **STO1** | **STO2** | **STO3** |
| Cheddar | $I_1^0 = 2.8$ | $I_2^0 = 3.12$ | $I_3^0 = 3.5$ |
| | $dI_1/dt = +1.5$ | $dI_2/dt = -0.07$ | $dI_3/dt = -0.8$ |
| Brie | $I_1^0 = 2.7$ | $I_2^0 = 2.7$ | $I_3^0 = 3.8$ |
| | $dI_1/dt = +7.2$ | $dI_2/dt = +3$ | $dI_3/dt = -6$ |
| Cheshire | $I_1^0 = 2.7$ | $I_2^0 = 3.3$ | $I_3^0 = 3.8$ |
| | $dI_1/dt = +3.6$ | $dI_2/dt = +0.72$ | $dI_3/dt = -6$ |
| Stilton | $I_1^0 = 4.0$ | $I_2^0 = 2.5$ | $I_3^0 = 5.4$ |
| | $dI_1/dt = +5.4$ | $dI_2/dt = +5.4$ | $dI_3/dt = -6$ |

**Extended data Figure 5.** Trained calibration parameters to classify each category of cheese in simulations.

Following the same procedure used in experiments, the simulated recognition performances are evaluated for each class of cheese using the associated initial conditions (Extended data Figure 5). The recognition rates obtained through this procedure are shown in Extended data Figure 6. Overall, the network responds correctly to 96% of the inputs.

| Cheese that the simulated oscillators are trained to detect | Presented cheese (10 datapoints) | | | | Correct response of the simulated network |
| --- | --- | --- | --- | --- | --- |
| | Cheddar | Brie | Cheshire | Stilton | |
| | Number of recognitions (out of 10) | | | | |
| **Cheddar** | 10 | 0 | 0 | 1 | **97.5 %** |
| **Brie** | 0 | 10 | 0 | 3 | **92.5 %** |
| **Cheshire** | 0 | 0 | 10 | 0 | **100 %** |
| **Stilton** | 1 | 0 | 0 | 9 | **95 %** |

**Extended data Figure 6.** Recognition rates obtained through the simulated network of coupled oscillators.